\documentclass[12pt]{article}
\usepackage{graphicx}
\begin{document}
\hyphenation{anti-fermion}
\baselineskip = 6.5mm
\topmargin= -15mm
\textheight= 230mm

\begin{center}
\begin{Large}
  {\bf Comment on "The massive Thirring model from XYZ spin chain" by Kolanovic et al. }
\end{Large}
\vspace{1cm}

Takehisa Fujita\footnote{fffujita@phys.cst.nihon-u.ac.jp},
Takuya Kobayashi\footnote{tkoba@phys.cst.nihon-u.ac.jp} and
Hidenori Takahashi\footnote{htaka@phys.ge.cst.nihon-u.ac.jp} $^*$

\vspace{5mm}
Department of Physics, Faculty of Science and Technology, \\ Nihon University, Tokyo, 
Japan

$ ^*$ Laboratory of Physics, Faculty of Science and Technology, \\ Nihon University, 
Chiba, Japan

\end{center}

\vspace{0.5cm}

\vspace{1cm}

\begin{center}

{\large ABSTRACT} 

\end{center}

It is shown that the continuum limit of the spin 1/2 Heisenberg XYZ model is 
far from sufficient for the site number of 16. Therefore, the energy spectrum 
of the XYZ model 
obtained by Kolanovic et al. has nothing to do with the massive Thirring model,  but 
it shows only the spectrum of the finite size effects.

\vspace{2cm}

The spin 1/2 Heisenberg model is a rich theory, which contains a variety
of field theories in the continuum limit. In particular, the Heisenberg
XYZ model is proved to be equivalent to the massive Thirring model in the
continuum limit.

Recently, Kolanovic, Pallua and Prester [1] solved the Heisenberg XYZ model
numerically to study the bound state problem in the massive Thirring model.
In their paper, they claim that the binding energies of the bosonic states
in the massive Thirring model are consistent with those of the semiclassical
calculations by Dashen et al.[2], contrary to the predictions by Fujita et al. [3-5].

In this comment, we show that the calculation of Kolanovic et al.[1] is
far from reliable due to the rough resolution of their calculations.
That is, the energy resolution in their calculation is not smaller
than the mass parameter they used, and therefore there is no chance
to obtain the binding energy of the system which should be in the continuum limit.
The spectrum they obtain is nothing but some factor times the resolution
${2\pi/{L}}$.

\vspace{1cm}

Now, the spin 1/2 Heisenberg XYZ model can be described by the following
hamiltonian,
$$ H=\sum_{i=1}^{N} \left( J_x S_i^x S_{i+1}^x + J_y S_i^y S_{i+1}^y +
J_z S_i^z S_{i+1}^z  \right) \eqno{(1)} $$
where $S_i^a$ is a spin operator at the site $i$. $J_x,J_y,J_z$ denote
the coupling constant, and $N$ is the site number.

According to Luther [6], this hamiltonian can be put into the following
equations of motion which describe the massive Thirring model,
$$  i\dot{ \psi}_1(k) = v_0k\psi_1(k)-im_0\psi^{\dagger}_2(-k)
-{4J_z\over L}\sum_{k'} \psi_1(k-k')\rho_2(k') \eqno{(2a)}  $$
$$  i\dot{ \psi}_2(k) = -v_0k\psi_2(k)+im_0\psi^{\dagger}_1(-k)
-{4J_z\over L}\sum_{k'} \psi_2(k-k')\rho_1(k') \eqno{(2b)}  $$
where the fermion mass $m_0$ and $v_0$ are defined as
$$ m_0={J_x-J_y\over{2a}}   \eqno{(3a)} $$
$$ v_0={1\over 2}{(J_x+J_y)}-{J_z\over{2\pi}} .  \eqno{(3b)} $$
Here,  $a$ denotes the lattice spacing constant, and
the box length $L$ is written as
$$ L=Na . \eqno{(4)} $$
The coupling constant $g$ is related to the  $J_x$ and $v_0$ as
$$ g= {2J_z\over{v_0}}.  \eqno{(5)} $$
Now, the value of $N$ in the calculation of ref.[1] is
$$ N=16 . $$
Since $\displaystyle{ L=(J_x-J_y) {N\over{2 m_0}} }$,
the resolution of the calculated spectrum becomes
$$ {2\pi\over{L}} =  {4\pi m_0\over{(J_x-J_y)N}} . \eqno{(6)}  $$
This is just comparable to the mass parameter $m_0$ in their calculations
as shown in Table 1.
Thus, it is impossible to extract any information on the bound state energy
in the massive Thirring model. In order to obtain some reasonable results
on the continuum version of the Heisenberg XYZ model, one has to satisfy the condition
$$ {2\pi\over L} << m_0 << {2\pi \over L}N . \eqno{(7)} $$
This suggests that, if one wants to obtain any reliable information
on the bound state of the massive Thirring model, one has to have the site number $N$
which is at least larger than $N=1000$.

\begin{table}
\caption{ The values of $m_0$ and $2\pi/L$ in units of $1/a$. 
In ref.[1], $N=16$ is taken.  }

\begin{center}
\begin{tabular}{c|c|c}\hline
$N$ & $2\pi/L$ &$m_0$\\ \hline\hline
{\bf 16} & 0.393 &  \\ 
100 & $6.28\times 10^{-2}$ & 0.1\\
1000 & $6.28\times 10^{-3}$ \\ \hline
\end{tabular}
\end{center}
\end{table}

In case  $N$ is small, then one obtains the spectrum
which is just some factor times the resolution $ {2\pi / L} $. 
This point is illustrated in figures 1 and 2. First, we show in fig.1 the 
calculation of the bound state spectrum in the massive Thirring model 
with the resolution of ${2\pi/{L}}$ [4]. There, one sees that the excited states 
above the free fermions are continuum states which are measured 
in units of $2\pi/L$. 
In fig. 2, we show the calculated results of the spectrum with $N=14$ site 
in the Heisenberg XYZ model. Even though their calculations 
are carried out with $N=16$,  there is practically no difference 
between $N=14$ and $N=16$ cases. 

Here, we should comment on the lattice calculations in general. In terms of 
the correlation length, one may calculate the ground state energy in 
the lattice calculations. In this case, the condition of the validity 
in the lattice calculations is that the correlation length must be 
much smaller than the box length, but  must be much larger than 
the lattice spacing. 

Here, however, one cannot discuss 
the energy difference between the lowest state and the excited states (including 
the continuum states ) because the energy difference between the lowest state and the 
excited states should be much larger than the resolution of $2\pi/L$. 
This is just the condition we imposed. 

Therefore, the results of ref.[1] can not be very reliable  
for the bound state spectrum of the massive Thirring model. 
But, of course, it does not necessarily mean that 
the spectrum of Dashen $et$ $al$. [2] is incorrect.

\vspace{2cm}
\underline{\bf References}
\vspace{0.5cm}

1. M. Kolanovic, S. Pallua and P. Prester, Phys. Rev. {\bf D62}, 025021 (2000)

2.  R. F. Dashen, B. Hasslacher and A. Neveu, Phys. Rev. {\bf D11}, 3432 (1975) 

3. T. Fujita and A. Ogura, Prog. Theor. Phys. {\bf 89}, 23 (1993) 

4. T. Fujita, Y. Sekiguchi and K. Yamamoto, Ann. Phys. {\bf 255}, 204 (1997) 

5. T. Fujita, T. Kake and H. Takahashi, Ann.  Phys. {\bf 282}, 100 (2000)

6. A. Luther, Phys. Rev. {\bf B14}, 2153 (1976)

\newpage

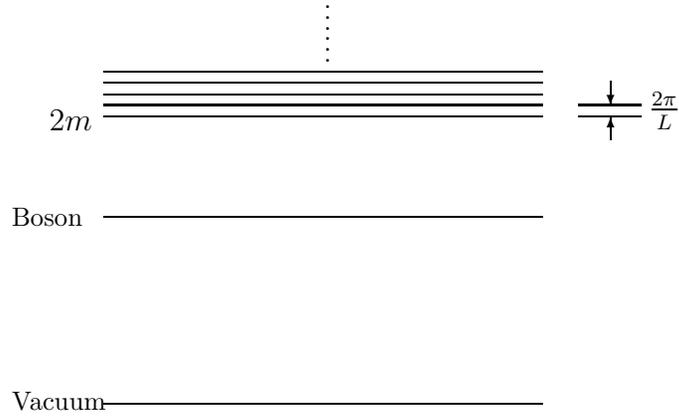
\begin{figure}
\setlength{\unitlength}{0.240900pt}
\ifx\plotpoint\undefined\newsavebox{\plotpoint}\fi
\begin{picture}(1500,900)(0,0)
\font\gnuplot=cmr10 at 10pt
\gnuplot
\sbox{\plotpoint}{\rule[-0.200pt]{0.400pt}{0.400pt}}%
\put(750,685){\makebox(0,0)[l]{\vdots}}
\put(319,561){\makebox(0,0)[l]{$2 m$}}
\put(260,120){\makebox(0,0)[l]{Vacuum}}
\put(260,409){\makebox(0,0)[l]{Boson}}
\put(750,735){\makebox(0,0)[l]{\vdots}}
\put(405,116){\usebox{\plotpoint}}
\put(405.0,116.0){\rule[-0.200pt]{165.980pt}{0.400pt}}
\put(405,409){\usebox{\plotpoint}}
\put(405.0,409.0){\rule[-0.200pt]{165.980pt}{0.400pt}}
\put(405,567){\usebox{\plotpoint}}
\put(405.0,567.0){\rule[-0.200pt]{165.980pt}{0.400pt}}
\put(405,585){\usebox{\plotpoint}}
\put(405.0,585.0){\rule[-0.200pt]{165.980pt}{0.400pt}}
\put(405,602){\usebox{\plotpoint}}
\put(405.0,602.0){\rule[-0.200pt]{165.980pt}{0.400pt}}
\put(405,620){\usebox{\plotpoint}}
\put(405.0,620.0){\rule[-0.200pt]{165.980pt}{0.400pt}}
\put(405,637){\usebox{\plotpoint}}
\put(405.0,637.0){\rule[-0.200pt]{165.980pt}{0.400pt}}
\multiput(1150,567)(10,0){10}{\line(1,0){8}}
\multiput(1150,585)(10,0){10}{\line(1,0){8}}
\put(1200,530){\vector(0,1){37}}
\put(1200,622){\vector(0,-1){37}}
\put(1260,565){$\frac{2 \pi}{L}$}
\end{picture}
\caption{A typical energy spectrum of the massive Thirring model \protect \newline 
\hspace{1.7cm} with the box length $L$. See ref.[4]. }
\end{figure}

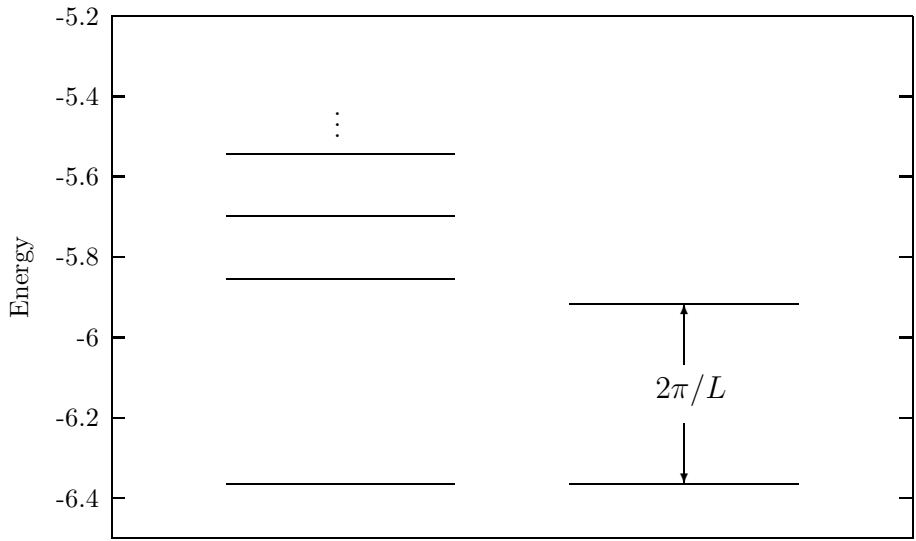
\begin{figure}
\vspace{0.5cm}
\setlength{\unitlength}{0.240900pt}
\ifx\plotpoint\undefined\newsavebox{\plotpoint}\fi
\sbox{\plotpoint}{\rule[-0.200pt]{0.400pt}{0.400pt}}%
\begin{picture}(1500,900)(0,0)
\font\gnuplot=cmr10 at 10pt
\gnuplot
\sbox{\plotpoint}{\rule[-0.200pt]{0.400pt}{0.400pt}}%
\put(181.0,103.0){\rule[-0.200pt]{4.818pt}{0.400pt}}
\put(161,103){\makebox(0,0)[r]{-6.4}}
\put(1419.0,103.0){\rule[-0.200pt]{4.818pt}{0.400pt}}
\put(181.0,229.0){\rule[-0.200pt]{4.818pt}{0.400pt}}
\put(161,229){\makebox(0,0)[r]{-6.2}}
\put(1419.0,229.0){\rule[-0.200pt]{4.818pt}{0.400pt}}
\put(181.0,355.0){\rule[-0.200pt]{4.818pt}{0.400pt}}
\put(161,355){\makebox(0,0)[r]{-6}}
\put(1419.0,355.0){\rule[-0.200pt]{4.818pt}{0.400pt}}
\put(181.0,482.0){\rule[-0.200pt]{4.818pt}{0.400pt}}
\put(161,482){\makebox(0,0)[r]{-5.8}}
\put(1419.0,482.0){\rule[-0.200pt]{4.818pt}{0.400pt}}
\put(181.0,608.0){\rule[-0.200pt]{4.818pt}{0.400pt}}
\put(161,608){\makebox(0,0)[r]{-5.6}}
\put(1419.0,608.0){\rule[-0.200pt]{4.818pt}{0.400pt}}
\put(181.0,734.0){\rule[-0.200pt]{4.818pt}{0.400pt}}
\put(161,734){\makebox(0,0)[r]{-5.4}}
\put(1419.0,734.0){\rule[-0.200pt]{4.818pt}{0.400pt}}
\put(181.0,860.0){\rule[-0.200pt]{4.818pt}{0.400pt}}
\put(161,860){\makebox(0,0)[r]{-5.2}}
\put(1419.0,860.0){\rule[-0.200pt]{4.818pt}{0.400pt}}
\put(181.0,40.0){\rule[-0.200pt]{303.052pt}{0.400pt}}
\put(1439.0,40.0){\rule[-0.200pt]{0.400pt}{197.538pt}}
\put(181.0,860.0){\rule[-0.200pt]{303.052pt}{0.400pt}}
\put(40,450){\makebox(0,0){\rotatebox{90}{Energy}}}
\put(1035,273){\makebox(0,0)[l]{$2 \pi/L$}}
\put(530,702){\makebox(0,0)[l]{\vdots}}
\put(181.0,40.0){\rule[-0.200pt]{0.400pt}{197.538pt}}
\put(1080,313){\vector(0,1){95}}
\put(1080,219){\vector(0,-1){94}}
\put(361,125){\usebox{\plotpoint}}
\put(361.0,125.0){\rule[-0.200pt]{86.483pt}{0.400pt}}
\put(361,447){\usebox{\plotpoint}}
\put(361.0,447.0){\rule[-0.200pt]{86.483pt}{0.400pt}}
\put(361,546){\usebox{\plotpoint}}
\put(361.0,546.0){\rule[-0.200pt]{86.483pt}{0.400pt}}
\put(361,644){\usebox{\plotpoint}}
\put(361.0,644.0){\rule[-0.200pt]{86.483pt}{0.400pt}}
\put(900,125){\usebox{\plotpoint}}
\put(900.0,125.0){\rule[-0.200pt]{86.483pt}{0.400pt}}
\put(900,408){\usebox{\plotpoint}}
\put(900.0,408.0){\rule[-0.200pt]{86.483pt}{0.400pt}}
\end{picture}
\caption{The energy spectrum of several lowest states \protect \newline
\hspace{1.7cm} in the Heisenberg XYZ model with $m_0=0.5,\;N=14, \protect \newline
\hspace{1.7cm} \;J_z=0.9,\; 2 \pi/L=0.4488$ 
in units of $1/a$. }
\end{figure}

\end{document}